\documentclass{ws-procs9x6}            
\begin{document}
\title{Viscous Cosmology}

\author{C. M. S. Barbosa, J. C. Fabris$^*$, O. F. Piattella, H. E. S. Velten and W. Zimdahl}

\address{Departamento de F\'{\i}sica, Universidade Federal do Esp\'{\i}rito Santo,\\
Vit\'oria, Esp\'{\i}rito Santo CEP29075-910, Brazil\\
$^*$E-mail: fabris@pq.cnpq.br\\
www.cosmo-ufes.org}

\begin{abstract}
We discuss the possibility to implement a viscous cosmological model, attributing to the dark matter component a behaviour described by bulk viscosity.
Since bulk viscosity implies negative pressure, this rises the possibility to unify the dark sector. At the same time, the presence of dissipative effects
may alleviate the so called small scale problems in the $\Lambda$CDM model. While the unified viscous description for the dark sector does not lead to consistent results,
the non-linear behaviour indeed improves the situation with respect to the standard cosmological model. 
\end{abstract}


\bodymatter

\section{Introduction}

The current standard cosmological model is very successful in describing many aspects of the observed universe \cite{mukha}. The spectrum of the anisotropies of the cosmic microwave background radiation (CMB), the large scale structure of matter distribution at linear level (LSS), the present stage of accelerated expansion of the universe, are some of the features of the observed universe very well described by the standard cosmological model. In order to incorporate successfully the description of the different observational data, two exotic components must be introduced in the cosmic budget, besides the ordinary components represented by baryons, radiation and neutrinos: dark matter and dark energy are necessary in order to explain structure formation and the present stage
of accelerated expansion. These components remain without any direct detection.

Even though, the standard cosmological model, dubbed $\Lambda$CDM model, has problems at non-linear level in the matter agglomeration \cite{wei}. Due to the fact that dark matter is a pressureless fluid, there is an excess of power at non-linear level, represented mainly by the prediction of an abundance of sub-structures around a galaxy about one order of magnitude higher than observations indicate. Moreover, the profile of the galactic density predicted by the $\Lambda$CDM model indicates a divergence (cusp) in its center, while observations reveal a core. Such discrepancies appear in many different analysis of
non-linear structure.

In the present text, we will develop one of the possibilities to cope with the problems of the $\Lambda$CDM model at small scales. It consists in considering that dark matter may behave as a viscous fluid. This possibility may lead to the avoidance of excess of power at galaxy and cluster of galaxy level.
In doing so, we may describe the viscous dark matter fluid using two different formalisms: the Eckart formalism, which supposes that the viscous fluid reaches
instantaneously the equilibrium at each momentum; the causal M\"uller-Israel-Stewart formalism, which includes a relaxation time. In both cases, there
are good results for some aspects of cosmology, but also important drawbacks.  We describe these different faces of the viscous cosmology model, and
point to some of possible solutions that may lead to a realistic model.

\section{The Different Viscous Formalisms}

General Relativity is defined by the following equations:
\begin{eqnarray}
R_{\mu\nu} - \frac{1}{2}g_{\mu\nu}R &=& 8\pi G T_{\mu\nu},\\
{T^{\mu\nu}}_{;\mu} &=& 0,
\end{eqnarray}
where $T^{\mu\nu} = (\rho + p)u^\mu u^\nu - pg^{\mu\nu}$.
Using the flat Friedmann-Lema\^{\i}tre-Robertson-Walker (FLRW),
\begin{eqnarray}
ds^2 = dt^2 - a(t)^2[dx^2 + dy^2 + dz^2],
\end{eqnarray}
those equations lead to the following equations of motion:
\begin{eqnarray}
\biggr(\frac{\dot a}{a}\biggl)^2 &=& \frac{8\pi G}{3}\rho, \\
2\frac{\ddot a}{a} + \biggr(\frac{\dot a}{a}\biggl)^2 &=& - 8\pi G p,\\
\dot\rho + 3 \frac{\dot a}{a}(\rho + p) &=& 0.
\end{eqnarray}

The $\Lambda$CDM model is based on the following identification:
\begin{eqnarray}
\rho = \rho_m + \rho_\Lambda, \quad p = p_m + p_\Lambda, \quad p_m = 0, \quad p_\Lambda = - \rho_\Lambda,
\end{eqnarray}
where $m$ and $\Lambda$ stands for the matter and dark energy components, respectively. Each component conserves separately.

The viscous cosmology is based in considering that the dark matter sector presents a dissipative behaviour represented by the bulk viscosity, such that
$p = p_v$, where $p_v$ stands for viscosity. There are two main descriptions of viscosity. The first one is the Eckart's formalism, that is non causal, given
by,
\begin{eqnarray}
p_v = - \xi(\rho)u^\mu_{;\mu},
\end{eqnarray}
where $\xi(\rho)$ is the bulk viscosity coefficient which depends on the fluid density. The viscous pressure is negative and it may contribute to the accelerated expansion of the universe. In fact, there is even the possibility that a viscous fluid may play the r\^ole of dark energy, unifying the dark sector.

The other formalism is the causal formalism of M\"uller-Israel-Stewart, defined by the transport equation,
\begin{eqnarray}
\label{mis}
\tau\Pi^{\bullet} + \Pi = -\theta\xi - \frac{1}{2}\tau\Pi\left[\theta - \frac{\left(\xi/\tau\right)^\bullet}{\left(\xi/\tau\right)} - \frac{T^\bullet}{T}\right]\;,
\end{eqnarray}
where we have defined $p_v \equiv \Pi$, and $\tau$ is the relaxation time. The superscript "$\bullet$" indicates a covariant time derivative, $u^\mu\partial_\mu$. When the relaxation time $\tau$ goes to zero, we recover the Eckart's formalism.

In what follows, for both formalisms, we will choose $\xi = \xi_0\rho^\nu$. We remark that for the Eckart's formalism such choice leads to a structure for
the background evolution very similar to the Generalised Chaplygin gas model (GCG) which is a kind of prototype of unification models for the dark sector.
The GCG model is characterised by the equation of state,
\begin{equation}
p = - \frac{A}{\rho^\alpha},
\end{equation}
where $A$ and $\alpha$ are - in principle - free parameters. The GCG gives very good results for the background evolution of the universe \cite{colistete}, but it faces
many problems at perturbative level \cite{zimdahl}.
As it will be seen below, at perturbative level, the GCG model and the viscous model are very different, what gives hopes that
the viscous model may cure the problems of the GCG model.

\section{Viscous Cosmology with the Eckart's Formalism}

Using the Eckart's formalism, the equations of motion read,
\begin{eqnarray}
\biggr(\frac{\dot a}{a}\biggl)^2 &=& \frac{8\pi G}{3}(\rho_v + \rho_b + \rho_\Lambda),\\
\dot\rho_v + 3\frac{\dot a}{a}\rho_v = 3\biggr(\frac{\dot a}{a}\biggl)^2\xi_0\rho_v^\nu&,& \quad \rho_b = \rho_{b0}a^{-3}, \quad \rho_{\Lambda} = \mbox{constant}.
\end{eqnarray}
In these expressions, the subscript $v$, $\Lambda$ and $b$ stand for {\it viscous}, {\it cosmological constant} and {\it baryonic} components respectively.

In the absence of the baryonic component, these equations are equivalent to those of the GCG model, with $\nu = - \alpha - \frac{1}{2}$.
Hence, the viscous model may be a candidate for the unified model for the dark sector. However, there are important tensions when this idea is implemented.
Using the supernova test (Union 2 sample) it comes out that the results for the viscous model are very competitive, as it happens for the GCG model. As in the GCG model, negative values of $\nu$ ($\alpha$) are preferred. However, quite large values of the parameter $\xi_0$ are necessary to fit the data.

The perturbative analysis in the case of the Eckart's formalism is more well posed than in the case of GCG. The reason is that in GCG model
imaginary sound velocities appear when $\alpha < 0$. In the viscous model, the perturbation of the pressure reads,
\begin{eqnarray}
\delta p &=& - 3H\xi'\delta\rho - \xi\biggr(\theta - \frac{\dot h}{2}\biggl). \nonumber
\end{eqnarray}
and are intrinsically non-adiabatic. This allows to consider any value of $\nu$.

At perturbative level the results are quite degenerate with respect to $\nu$, but in order to reproduce the power spectrum data (2dFRGS data), very small values of $\xi_0$ are preferred \cite{carol}. Figure 1, the two upper panels, illustrate this tension. Hence, in the unified model version of the viscous cosmology it is not possible to fit simultaneously the
background and perturbative data using the Eckart's formalism.

Such tension can be relaxed if a cosmological constant is considered. Even though a fractional density of order of $\Omega_{\Lambda} \sim 0.7$ is necessary, see figure 1, two lower panels. Hence, the model is reduced to the usual standard one but considering a viscous dark matter fluid. Even if the unified scenario
is abandoned, good results are obtained at non-linear level: if the problems of excess of power are not completely solved, they are at least considerably alleviated. The non-linear results using the spherical collapse model described in reference \cite{velten}
indicate that a viscous coefficient of order of $\bar\xi_0 \sim 10^{-8}$ (in convenient unities, see reference \cite{velten}) would lead to almost the same predictions
as the standard $\Lambda$CDM model but with a slightly suppressed
growth of the smallest cosmological structures in the universe. For this value of the viscosity coefficient, for example, the abundance of number counts with masses of
order $M \sim 10^9 M_\odot$ is reduced by almost one order of magnitude.

\begin{figure}[!t]
\begin{minipage}[t]{0.4\linewidth}
\includegraphics[width=\linewidth]{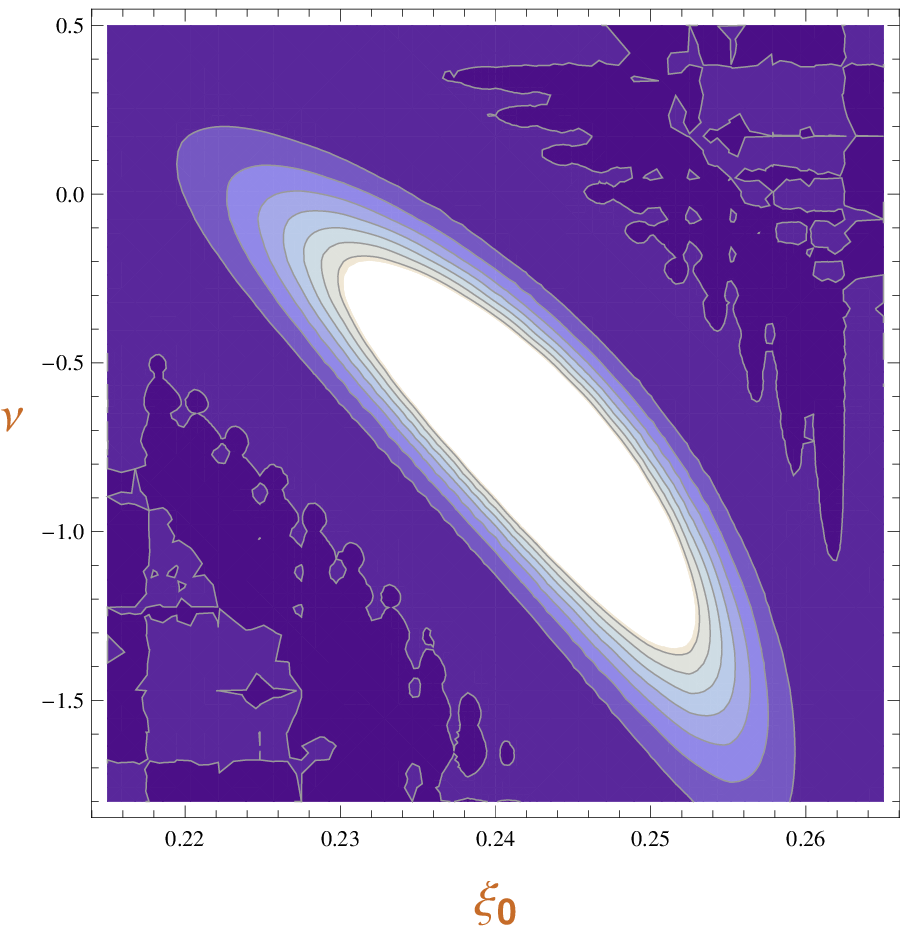}
\end{minipage} \hfill
\begin{minipage}[t]{0.4\linewidth}
\includegraphics[width=\linewidth]{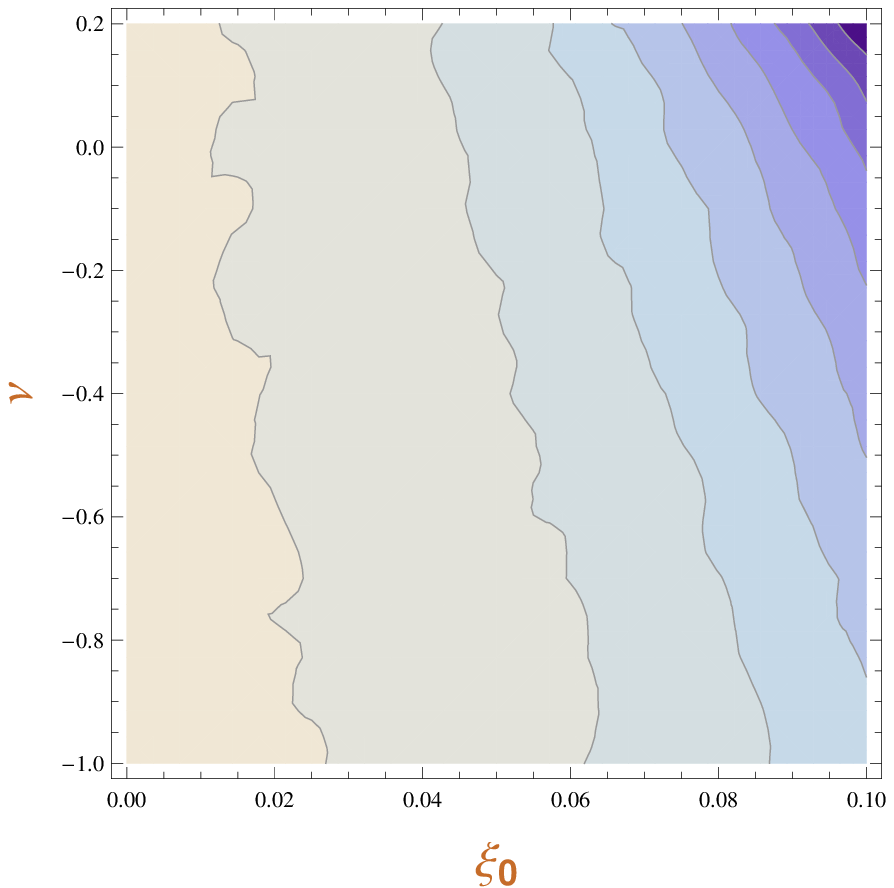}
\end{minipage} \hfill
\begin{minipage}[t]{0.4\linewidth}
\includegraphics[width=\linewidth]{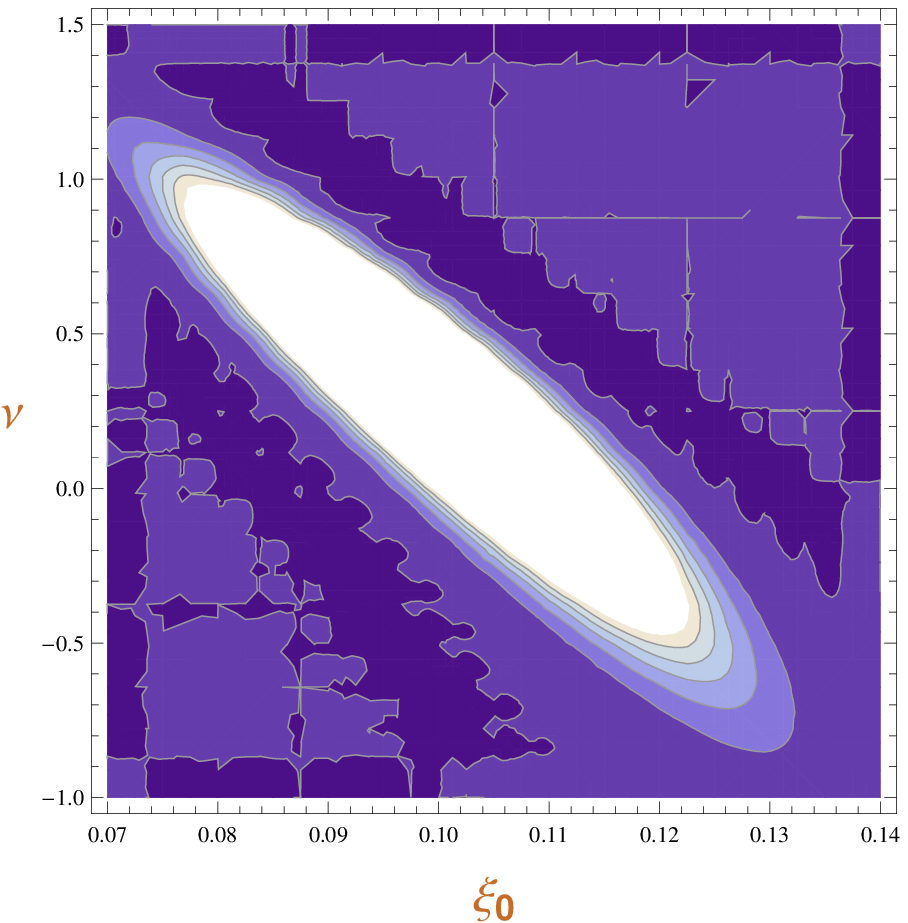}
\end{minipage} \hfill
\begin{minipage}[t]{0.4\linewidth}
\includegraphics[width=\linewidth]{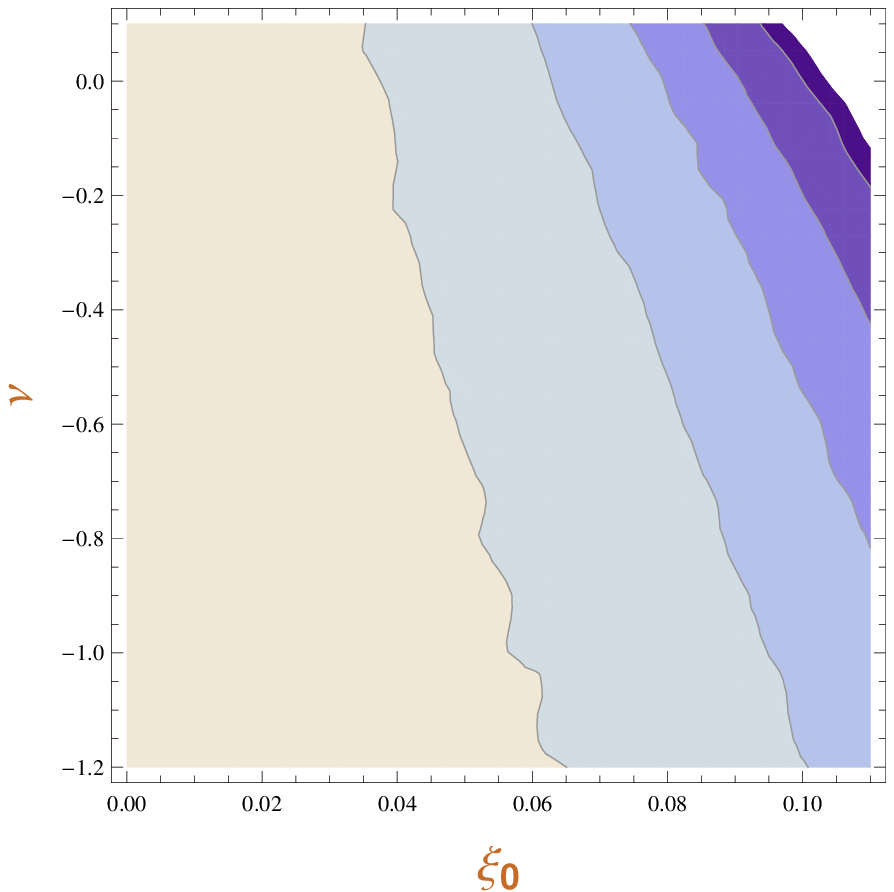}
\end{minipage} \hfill
\caption{Two-dimensional probability distribution function (PDF) using SN Ia data (left) and 2dFGRS power spectrum (right) with $\Omega_\Lambda = 0$ (upper panels) and $\Omega_\Lambda = 0.7$ (lower panels).}
\end{figure}

\section{Viscous Cosmology in the MIS Formalism}

The employment of the causal formalism of M\"uller-Israel-Stewart (MIS) may change the scenario described before for the viscous cosmology in the Eckart's formalism. In some sense, this formalism is more physical, since the relaxation time introduced in the formalism avoids the causality problems of the Eckart's formalism. But, on the other hand, the MIS formalism introduces new parameters, like the relaxation time itself. In the absence of a fundamental microphysical model for the viscous fluid, these parameters must be fixed phenomenologically. 

One possibility for the relaxation time is to consider it as proportional to the Hubble parameter that is the only time scale available. Moreover, it is possible to consider the truncated form of the MIS formalism, given by,
\begin{eqnarray}
\label{mis}
\tau\Pi^{\bullet} + \Pi = -\theta\xi \;.
\end{eqnarray}
The results will not change dramatically if we use the complete MIS formalism or its truncated version.

The situation for the viscous cosmological model is, generally, more dramatic if the Integrated Sachs-Wolfe (ISW) effect is taken into account. Figure 2 shows the evolution of the gravitational potential in the viscous model compared with the $\Lambda$CDM results. The computation shows an important discrepancy, what leaves the viscous model, in the Eckart and MIS formalism with difficulties to explain the plateau of the CMB spectrum \cite{v-z}. Such problems have already been remarked also in reference \cite{barrow}. However, these results were obtained in the context of the unified model, with zero cosmological constant. The results may change significantly if a cosmological term is introduced.

\begin{center}
\begin{figure}[!t]
\begin{minipage}[t]{0.45\linewidth}
\includegraphics[width=\linewidth]{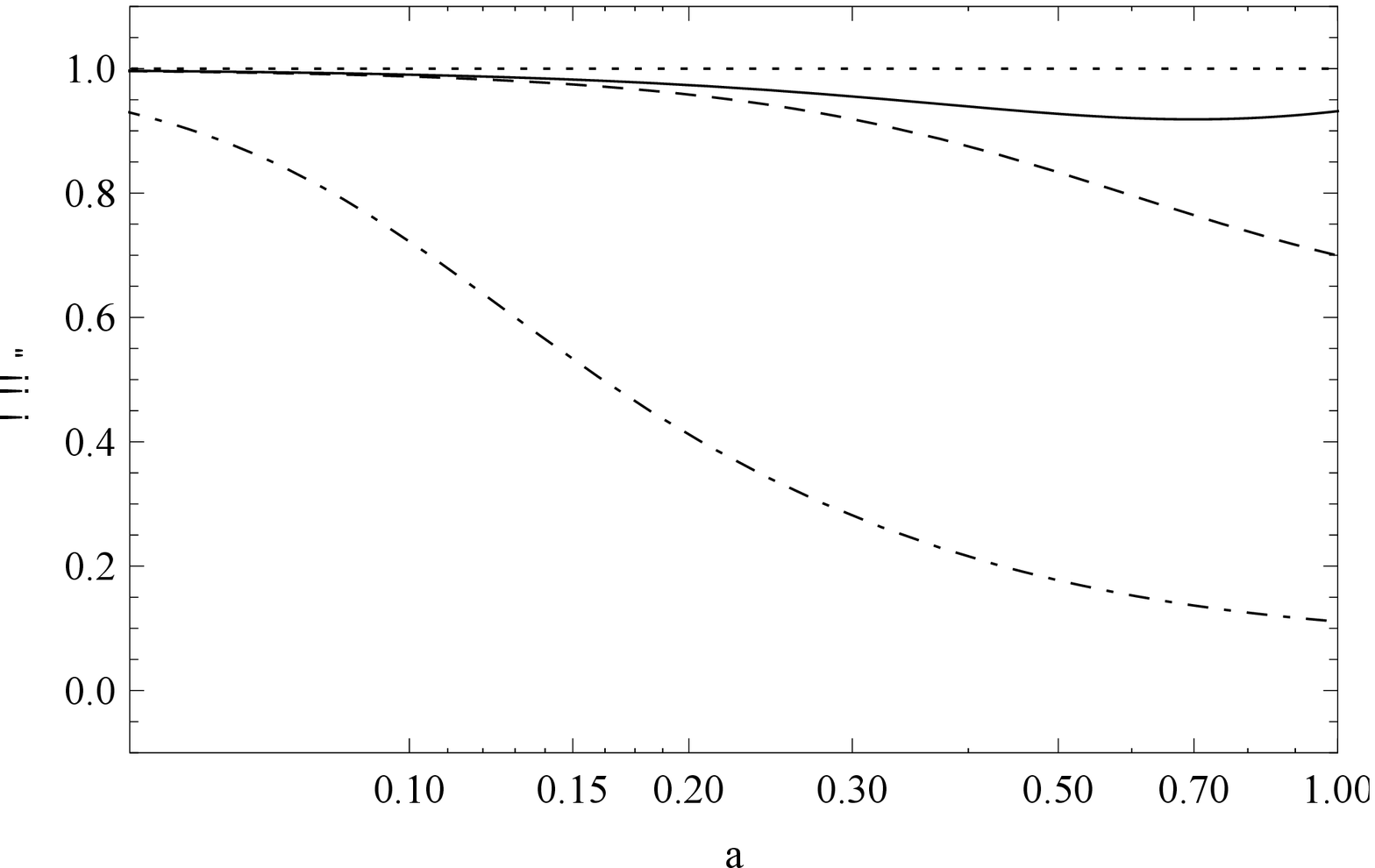}
\end{minipage} \hfill
\begin{minipage}[t]{0.45\linewidth}
\includegraphics[width=\linewidth]{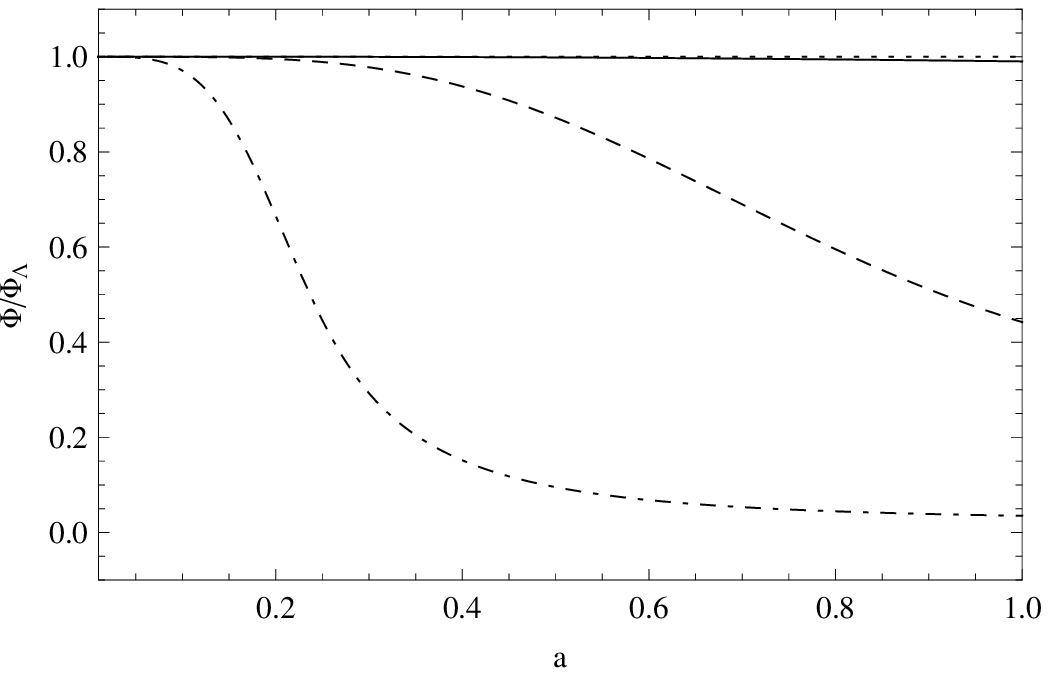}
\end{minipage} \hfill
\caption{The ratio of the evolution of the gravitational potential in the Eckart's formalism (right panel) and truncated MIS formalism, with respect to the $\Lambda$CDM case, for some different perturbative modes characterised by the wavenumber $k$, with $k = 10^{-4}h
, 10^{-3} h
, 10^{-2} h\,
Mpc^{-1}$
(solid, dashed, dot-dashed lines, respectively). See reference \cite{v-z} for details.}
\end{figure}
\end{center}

\section{Conclusions}

We have described in the present text the main proposals of a viscous cosmological model. Two main formalisms have been used: the Eckart's non-causal formalism and the M\"uller-Israel-Stewart formalism. There are two main goals in introducing such viscous model for the dark sector. The first one is to cope with the problems of the $\Lambda$CDM model at small scales, that constitutes a considerable challenge for the standard cosmological model. The second one is the possibility to unify the two components of the dark sector into a single component in a spirit similar to the GCG model.

In general, good results are obtained for the background description of the evolution of the universe. In particular, using the SN Ia data, the viscous models lead to a very good fitting. However, the situation becomes more involved at perturbative level. First of all, there is a tension in the parameter estimation using perturbations and background evolution. Such discrepancy can be alleviated if a cosmological term is introduced. Hence, it seems that the unified model for
the dark sector is not a viable route in this viscous cosmology program.

Moreover, the Integrated Sachs-Wolfe effect reveals an important suppression of power with respect to the $\Lambda$CDM model. This implies that there is no possibility in general to fit conveniently the CMB spectrum. Again the introduction of a cosmological term may change the  situation leading to a better fitting.

In any case, even if the unified model is abandoned, there are very good results at non-linear level, where the problems of the $\Lambda$CDM model are
considerably attenuated. This viscous model with cosmological constant is dubbed $\Lambda v$CDM and it is free from any extra drawback. This motivates the search for more realistic and sophisticated viscous cosmological models.

\section*{Acknowledgments} We thank CNPq (Brazil) and FAPES (Brazil) for partial financial support.

\end{document}